\documentclass[a4paper,11pt]{article}
\usepackage{jheppub}
\usepackage{amsmath}
\usepackage{amsthm}
\usepackage{amsfonts}
\usepackage{amssymb}
\usepackage{textcomp}
\usepackage{graphicx}
\usepackage{bm}
\usepackage{color}
\usepackage{verbatim}
\usepackage{subcaption}
\usepackage{graphicx,color}
\usepackage{epsfig}

\usepackage{amsfonts,amsmath,amssymb}

\newcommand{\be}{\begin{equation}}
\newcommand{\ee}{\end{equation}}
\newcommand{\bea}{\begin{eqnarray}}
\newcommand{\eea}{\end{eqnarray}}
\numberwithin{equation}{section}

\begin{document}

 \title{Large D holography with metric deformations}

 \author[a]{Tomas Andrade}
 \author[b]{Christiana Pantelidou}
 \author[c]{Benjamin Withers}

 \affiliation[a]{Departament de Fisica Quantica i Astrofisica \& Institut de Ciencies del Cosmos (ICC)\\ Universitat de Barcelona, Marti i Franques 1, 08028 Barcelona, Spain} 
 \affiliation[b]{Centre for Particle Theory and Department of Mathematical Sciences,\\ Durham University
Durham, DH1 3LE, U.K.} 
 \affiliation[c]{Department of Theoretical Physics, University of Geneva,\\ 24 quai Ernest-Ansermet, 1214 Geneve 4, Switzerland}


\abstract{We consider Einstein gravity in AdS in the presence of a deformed conformal boundary metric, in the limit of large spacetime dimension. At leading order we find a new set of effective near-horizon equations. These can be understood as covariant generalisations of the undeformed equations with new source terms due to the curvature. We show that these equations are given by the conservation of the exact second-order Landau-frame hydrodynamic stress tensor. No derivative expansions are invoked in this identification. We use the new equations to study CFTs with 2d lattice deformations, computing their quasi-normal mode spectra and thermal conductivities, both numerically and analytically to quartic order in small lattice amplitude. Many of our results also apply to asymptotically flat spacetimes.}

\maketitle

\section{Introduction}

Understanding strongly interacting quantum systems presents a significant theoretical challenge.
A promising step forward is provided by holography \cite{Maldacena:1997re} relating certain strongly coupled CFTs in $d$ dimensions to Einstein gravity in asymptotically AdS$_{d+1}$ spacetimes. This duality can be used as a tool to gain new insight into strongly interacting systems, however to do so we must tackle the appropriate set of problems in general relativity. Analytic solutions of the Einstein equations are rare, and typically correspond to highly symmetric states. Therefore what remains is a technical challenge to construct and understand solutions of the Einstein equations with no particular symmetries. One approach is to turn to numerics, and whilst this approach can be very effective when there is a sharp question to answer or specific quantity to calculate, it is perhaps less so when seeking to obtain new insight. Any analytical handle we can gain in this context is therefore valuable.

In this paper we turn to the large-dimension ($d$) expansion of general relativity to gain such a handle. By treating $d$ as a free parameter we can use it as a natural way to do perturbation theory. One simple point to expand around is $d\to\infty$, since for a black hole (with an appropriate scaling of parameters) nontrivial gravitational fields are localised within a distance $1/d$ from the horizon \cite{Asnin:2007rw, Emparan:2013moa, Emparan:2013xia}. This leads to an effective reduction in the dimension of the theory governing the dynamics of black holes, described only by equations that depend on the directions on the horizon \cite{Emparan:2015hwa, Bhattacharyya:2015dva, Suzuki:2015iha, Suzuki:2015axa, Emparan:2015gva}. The large-$d$ expansion may then be used to quantitatively understand systems at say, $d=2,3,4$ by studying the convergence of systematic corrections in $1/d$, or in a more conservative approach, used to understand gravity and strongly interacting systems on more qualitative grounds without reference to a specific dimension. 

In the context of holography large-$d$ techniques have been used to study a wide variety of physical situations including superfluids \cite{Emparan:2013oza, Romero-Bermudez:2015bma}, quasinormal modes \cite{Emparan:2015rva}, momentum relaxation and transport \cite{Andrade:2015hpa}, a Riemann problem \cite{Herzog:2016hob}, and turbulence \cite{Rozali:2017bll}. Polarised black holes arising from chemical potential deformations were considered in \cite{Emparan:2016sjk, Iizuka:2018zgt}. More widely there have been a number of interesting extensions and applications of the large-$d$ formalism \cite{Tanabe:2015hda, Bhattacharyya:2015fdk, Emparan:2016sjk, Tanabe:2016opw, Rozali:2016yhw, Dandekar:2016fvw, Dandekar:2016jrp, Bhattacharyya:2016nhn,  Chen:2017wpf, Chen:2017hwm, Bhattacharyya:2017hpj, Miyamoto:2017ozn, Chen:2017rxa,  Herzog:2017qwp, Dandekar:2017aiv, Emparan:2018bmi}. 

In this paper we use the large-$d$ expansion to study CFTs on inhomogeneous and time dependent background spacetimes, with metric $h_{AB}(t,\vec{x})$. In detail, we consider non-trivial dependence on $p$ of the $d-1$ field theory spatial directions as follows,
\be
h_{AB}(t,\vec{x}) dx^A dx^B = -\left(1 - \frac{\gamma_{tt}(t,\vec{x})}{n}\right)dt^2 - \frac{2}{n} \zeta_i(t,\vec{x}) dt dx^i+ \frac{1}{n}\gamma_{ij}(t,\vec{x})dx^idx^j + \frac{1}{n}d\vec{y}^2,  \label{bdymet}
\ee
where $\vec{x}$ labels the $p$ field theory spatial directions, leaving a large number $n= d-p-1$ of spatial directions labelled by $\vec{y}$ in which we retain $ISO(n)$ symmetry. Physically we may use such deformations to model features of strongly interacting systems. For example, by using $\gamma_{ij}(t,\vec{x})$ we may model strain disorder.\footnote{For example, as studied in the hydrodynamic limit by \cite{Scopelliti:2017sga}.} By turning on $\zeta_i(t,\vec{x})$ we may introduce a time-dependent thermal gradient, which for example, allows us to compute AC thermal conductivities by measuring the resulting heat current. 

A key result of our work is a new set of effective near-horizon equations at leading order in $1/d$ which includes these metric deformations. We show that these equations are related to the equations in the absence of such deformations \cite{Emparan:2016sjk} by covariantising them with respect to $\gamma_{ij}$ and by adding source terms provided by gradients of $\gamma_{ij},\zeta_i, \gamma_{tt}$. In addition to asymptotically locally AdS spacetimes our results also describe asymptotically flat spacetimes through the AdS-Ricci flat correspondence \cite{Caldarelli:2012hy,Caldarelli:2013aaa}. These equations are presented in \eqref{resultsnozeta0} and \eqref{resultsnozetai}.

A second key result of our work is the relation to hydrodynamics. We show that the near-horizon effective equations are given exactly by the covariant conservation of Landau-frame relativistic conformal hydrodynamics on the curved background \eqref{bdymet}, truncated to second order in derivatives. To be precise, to leading order in large-$d$ the gravitational equations are given simply by the large-$d$ limit of the covariant conservation equation,
\be
\nabla_AT^{AB}=0 \label{cons}
\ee
where
\be
T^{AB} \equiv \epsilon\, U^A U^B + \frac{\epsilon}{d-1} \Delta^{AB} -\eta\, \sigma^{AB} + c_1 \mathcal{O}^{AB}_1, \label{Thydro}
\ee
where $U^A$ is a unit-normed timelike vector, $\Delta^{AB}$ the projector orthogonal to $U^A$, $\sigma^{AB}$ is the associated shear tensor and $\mathcal{O}^{AB}_1$ is a term second order in derivatives as defined in \cite{Baier:2007ix}, given later in \eqref{Odefs}. The expressions \eqref{cons} and \eqref{Thydro} here are covariant with respect to the CFT metric $h_{AB}$. Specifically, we focus on the case of interest for condensed matter applications, $p=2$, i.e. non-trivial dynamics in two of the spatial directions. We find the following transport coefficients for large-$d$ gravity at leading order 
\be
\eta = \frac{1}{4\pi}s, \qquad  c_1 \equiv \eta \tau_\Pi=  \frac{1}{16\pi^2} \frac{s}{T}.
\ee
We stress that no derivative expansion is invoked in going from \eqref{cons} and \eqref{Thydro} to the gravitational equations. Crucially for this to work the stress tensor \eqref{Thydro}, whilst second order and in Landau frame, is not of the usual BRSSS form \cite{Baier:2007ix}. In \cite{Baier:2007ix} the gradient expansion is reorganised at second and higher orders by invoking the lower-order equations of motion.  In doing so, ${\cal O}_1^{AB}$ is replaced by a collection of other terms at second order and at higher orders too. This reorganised expansion does not match the large-$d$ gravity equations unless one further invokes a derivative expansion. 
However here we emphasise that the near-horizon effective equations are given by the covariant conservation of the stress tensor \eqref{Thydro} without invoking a derivative expansion, and that is because we use the second order term ${\cal O}_1^{AB}$.\footnote{In previous work \cite{Rozali:2017bll} the use the BRSSS form at second-order has led to identification of large-$d$ gravity equations with hydrodynamics that is only approximate in the derivative expansion.}

As an application of our results we analyse solutions with explicit lattice sources. In the case of $p=2$ we deform 
the CFT with a 2 dimensional monochromatic lattice,
\be
\label{lattice def}
\gamma_{ij}(t,x,y) = \left(1 + A_0 \cos k_Lx \cos k_Ly\right) \delta_{ij}
\ee
where $x=x^1, y =x^2$, and solve the resulting near-horizon equations to construct a lattice black hole. In addition we compute the QNM spectra for these black holes, as well as their AC thermal conductivities, $\kappa^{ij}(\omega)$. We proceed first by perturbation theory order-by-order in small lattice amplitude, $A_0$, where we work to order $(A_0)^4$. We then turn to a full numerical treatment of our equations. 

Our results fulfil the expectations of a hydrodynamic system in the presence of translational symmetry breaking terms. 
At $A_0 = 0$, the modes of the system are given by shear and sound modes with the standard gapless dispersion relations. As a result, the  conductivity of the system has a delta 
function at zero frequency.
Turning on a non-zero lattice removes the delta function in the conductivity. Specifically, we find that at small $\omega\sim q\sim O(A_0)^2$ the longitudinal spatially-resolved conductivity is given by
\begin{equation}
\kappa^{xx}(\omega,q) = \frac{i \omega \alpha_1}{i \omega \left( - i \omega + \Gamma + q^2 \alpha_2 \right) - q^2 \alpha_3}+ O(A_0)^1\label{kappaintro}
\end{equation}
\noindent where $\Gamma$, $\alpha_i$ are parameters which depend on $k_L$ and $A_0$, which we give analytically in the case of perturbation theory in $A_0$. At leading order this expression scales like $(A_0)^{-2}$ diverging in the translationally invariant limit. Moreover when $q\sim O(A_0)^2$ there is a collision between the poles resulting in diffusive modes at long wavelengths. This is also reflected in the sound channel QNM spectrum which we compute. We note that this formula was also seen in a hydrodynamic-like theory modified by a momentum dissipation term in \cite{Davison:2014lua}. Similarly in the transverse channel we find an expression whose poles are given by the shear-channel QNMs,
\be
\kappa^{yy}(\omega,q) = \frac{i \alpha_1}{\omega + i \Gamma + i q^2}+ O(A_0)^1.
\ee

The rest of the paper is structured as follows. In section \ref{sec:gravity} we derive the new set of near-horizon effective equations in the presence of the curved, time dependent boundary metric \eqref{bdymet}. We then show in section \ref{sec:hydro} how the gravity variables are related to hydrodynamic Landau-frame variables, and demonstrate that the equations of motion follow exactly from the conservation of \eqref{Thydro}. In section \ref{sec:transport} we turn to our application: computing lattice solutions, their QNM spectra and AC thermal conductivity. We finish with a discussion of our results in section \ref{sec:discussion}.

\section{Black branes with $p=2$ in the large-d limit}\label{sec:gravity} 

We consider a $(d+1)$-dimensional theory of gravity with a  negative cosmological constant, described by the Lagrangian
\be
S_{up} = \int d^{d+1}X \sqrt{-G}\left(R(G)-2\Lambda \right) \label{upstairs}\,,
\ee
with $\Lambda = -\frac{1}{2}d(d-1)$. We adopt the following ansatz
\bea
G_{\alpha\beta}dX^\alpha dX^\beta &=& g_{ab} dx^a dx^b + e^{2\beta\phi(x)} \delta_{mn} dy^m dy^n\label{reduction}
\eea
where there are $n$ coordinates $y^m$ and $p+2$ coordinates $x^a$, so that $d = p + n + 1$. A convenient way to proceed is to dimensionally reduce on an n-torus parametrised by the $y^m$ coordinates and obtain theories of gravity in the reduced finite-dimensional spacetime with a dilaton field for the size of the compactified space. With the choice of $\beta = -p\alpha/n$, the equations of motion for the reduced theory become
\bea
R_{ab}+\alpha p \nabla_a \partial_b \phi -\frac{\alpha^2 p^2}{n}\partial_a\phi\partial_b\phi-\frac{2\Lambda}{n+p} g_{ab}  &=& 0\nonumber\\
\Box \phi - \alpha p (\partial \phi)^2 - \frac{2 \Lambda n}{\alpha p (n+p)} &=& 0
\eea
where $\alpha$ is an arbitrary $\phi$-normalisation.

Following \cite{Emparan:2016sjk, Rozali:2017bll} we adopt a Bondi-type ansatz for neutral branes, with the coordinate $R=r^n$,
\be
G_{\alpha\beta}dX^\alpha dX^\beta = 2 dt dr +r^2\left(-2 A dt^2 - 2F_i dt dx^i + G_{ij}dx^idx^j\right)+ r^2 G_\perp d\vec{y}^2
\ee
where the functions $A,F_i,G_{ij},G_\perp$ are functions of $\{t,R,x^i\}$, with $i=1,2$.
We shall use the non-Einstein frame reduction, as in \eqref{reduction}. For consistency with \eqref{reduction} we must identify the dilaton with
\be
e^{2\beta\phi} = r^2 G_\perp(t,R,x^k)
\ee
together with the metric,
\be
g_{ab}dx^a dx^b = 2 \frac{R^{\frac{1}{n}-1}}{n}dt dR +R^{\frac{2}{n}}\left(-2 A dt^2 - 2F_i dt dx^i + G_{ij}dx^idx^j\right).
\ee
Next we solve the equations order-by-order in an expansion in $1/n$ for the functions $A,F_i,G_{ij},G_\perp$. For the leading terms we adopt,
\bea
G_\perp &=& \frac{1}{n} + G_\perp^{(3)}\frac{1}{n^3}+\ldots\\
A &=& \left(\frac{1}{2} - \frac{a(t,x,y)}{2R}\right) - \frac{\gamma_{tt}(t,x,y)}{2n} +  A^{(1)} \frac{1}{n} + \ldots\\
F_i &=& \left(\zeta_i(t,x,y) + \frac{p_i(t,x,y)}{R}\right)\frac{1}{n} + F_i^{(2)}\frac{1}{n^2}  + \ldots\\
G_{ij} &=& \gamma_{ij}(t,x,y)\frac{1}{n}  + G_{ij}^{(2)}\frac{1}{n^2}  + \ldots
\eea
We have here chosen to extend to the case of a boundary metric deformation in the form of $\gamma_{ij}(t,x,y)$, 
$\gamma_{tt}(t,x,y)$ and $\zeta_i(t,x,y)$.
Also $G_\perp^{(3)}, A^{(1)},F_i^{(2)},G_{ij}^{(2)}$ have been determined as functions of the lower order data but are omitted from this presentation. The ellipses denote higher order parts of the expansion.
The complete boundary metric is given in \eqref{bdymet}.

With this expansion, subject to the condition that the solution is regular and that higher order pieces do not further adjust the boundary metric, we arrive at the following constraints on the data. For $\gamma_{ij} = \delta_{ij}, \zeta_i = 0,\gamma_{tt}=0$ we have exactly the equations as given in \cite{Emparan:2016sjk, Rozali:2017bll}, namely,
\bea
\partial_t a - \partial_i\partial^ia &=& -\partial_ip^i\\
\partial_t p_i - \partial_j\partial^j p_i &=& -\lambda\, \partial_i a - \partial_j\left(\frac{p_i p^j}{a}\right).
\eea
where indices are raised here with $\delta^{ij}$.
For $\gamma_{ij} \neq \delta_{ij},\zeta_i \neq 0,\gamma_{tt}\neq0$ we have terms which are the naive covariantisation of the above, plus sources due to the boundary deformations,
\bea
(\partial_t +K- \nabla_i\nabla^i)a &=& -\nabla_i\tilde{p}^i\,,\label{resultsnozeta0}\\
(\partial_t +K- \nabla_j\nabla^j )\tilde{p}_i  &=& -\lambda \,\nabla_i a - \nabla_j\left(\frac{\tilde{p}_i \tilde{p}^j}{a}\right) + \frac{\tilde{p}_i}{2}{\cal R} -\frac{\nabla_i(a^2 {\cal R})}{2a}- a \partial_iK + 2\nabla_m(a K^m_{~~i})\nonumber\\
&+& a \partial_t \zeta_i + 2 (\tilde{p}^j - \nabla^j a)\nabla_{[j}\zeta_{i]}+ \frac{a}{2}\nabla_i \gamma_{tt},\label{resultsnozetai}
\eea
where $\tilde{p} = p + a \zeta$. Here covariance is with respect to $\gamma_{ij}$, ${\cal R}$ is the Ricci scalar of $\gamma_{ij}$ and $K_{ij} = \frac{1}{2}\partial_t\gamma_{ij}$ with $K=\gamma^{ij}K_{ij}$. 

Given the above solution, the AdS-Ricci flat correspondence \cite{Caldarelli:2012hy,Caldarelli:2013aaa} can be used to generate an asymptotically flat solution, corresponding to the ansatz
\bea
ds^2=2dt\, dr-2Adt^2 -2\,F_i d\sigma^i dt+G_{i j}dx^idx^j+r^2 d\Omega_{n+1}^2
\eea
Equations \eqref{resultsnozeta0} and \eqref{resultsnozetai} apply for both the asymptotically flat and the asymptotically AdS case, with 
\begin{equation}
\lambda=
\left\{
	\begin{array}{ll}
		-1  & \mbox{for AF}\\
		+1 & \mbox{for AdS}.
	\end{array}
\right.
\end{equation}

\section{Large-d hydrodynamics\label{sec:hydro}}
In this section we start with conformal relativistic hydrodynamics in $d = n+p+1$ dimensions with a metric $h_{AB}$. First let us define the projector,
\be
\Delta^{AB} \equiv h^{AB} + U^{A} U^{B},
\ee
as well as the shear and vorticity tensors,
\bea
\sigma^{AB} &\equiv& \Delta^{A}_{\phantom{A}C}\Delta^{B}_{\phantom{B}D}\left(\nabla^CU^D + \nabla^D U^C - \frac{2}{d-1}h^{CD}\nabla\cdot U\right)\\
\omega^{AB} &\equiv& \frac{1}{2}\Delta^{A}_{\phantom{A}C}\Delta^{B}_{\phantom{B}D}\left(\nabla^CU^D - \nabla^D U^C\right).
\eea
With these definitions, the constitutive relations of conformal relativistic hydrodynamics in Landau-frame, truncated to second order in derivatives can be written \cite{Baier:2007ix}
\bea
T^{AB} &=& T^{AB}_{(0)} + T^{AB}_{(1)} + T^{AB}_{(2)},\\
T^{AB}_{(0)} &=& \epsilon\, U^A U^B + \frac{\epsilon}{d-1} \Delta^{AB},\label{eq:T0_def}\\
T^{AB}_{(1)} &= & -\eta\, \sigma^{AB},\\
T^{AB}_{(2)} &= &c_1 \mathcal{O}^{AB}_1+c_2 \mathcal{O}^{AB}_2+c_3 \mathcal{O}^{AB}_3+c_4 \mathcal{O}^{AB}_4+c_5 \mathcal{O}^{AB}_5\,,
\eea
where the terms in the second order stress tensor are given by 
\begin{align}
\mathcal{O}^{AB}_1 &=(\Delta^{AC}\Delta^{BD}-\frac{\Delta^{AB}}{d-1}\Delta^{CD}) \left(R_{CD}-(d-2)\nabla_C\nabla_D  \ln{T}+(d-2)\nabla_C  \ln{T}\, \nabla_D \ln{T} \right),\,\nonumber\\
\mathcal{O}^{AB}_2 &=(\Delta^{AC}\Delta^{BD}-\frac{\Delta^{AB}}{d-1}\Delta^{CD})\left(R_{CD}-(d-2) U^MU^N R_{MCDN}\right),\,\nonumber\\
\mathcal{O}^{AB}_3 &=\sigma^A_{~C}\,\sigma^{BC}-\frac{\Delta^{AB}}{d-1}\sigma^{CD}\sigma_{CD},\,\nonumber\\
\mathcal{O}^{AB}_4 &=\sigma^A_{~C}\,\omega^{BC} +\omega^A_{~C}\,\sigma^{BC}, \,\nonumber\\
\mathcal{O}^{AB}_5 &=\omega^A_{~C}\,\omega^{BC}-\frac{\Delta^{AB}}{d-1}\omega^{CD}\omega_{CD}.
\label{Odefs}
\end{align}
Here $R_{MCND}$ is the Riemann tensor of $h_{AB}$.
The hydrodynamic coefficients at second order, $c_i$, are related to the transport coefficients in the BRSSS formulation as follows
\begin{align}\label{eq:traspcoeffs}
\lambda_1 & = \frac{c_1}{2}+c_3\,\nonumber\\
\lambda_2 &= c_4\,\nonumber\\
\lambda_3 &= -2 c_1+c_5\,\nonumber\\
\eta \,\tau_{\pi} & =  c_1\,\nonumber\\
\kappa & =  c_1+c_2\,.
\end{align}

Next we turn to evaluating the hydrodynamic conservation equations in the large $n$ limit. We first identify $h_{AB}$ with our choice of deformed boundary metric \eqref{bdymet}, and focus on the cases of trivial dynamics in the large-$n$ directions, $U_m = \partial_m \epsilon = 0$, labelling $u_\mu = U_\mu$. Next the $(p+1)$-vector $u^\mu$, once unit-normed timelike and future-directed, can be written,
\be
u^\mu = \frac{1}{\sqrt{1-\frac{1}{n}\gamma_{tt}-\frac{1}{n}\gamma_{ij}\beta^i\beta^j+ \frac{2}{n}\zeta_i \beta^i}}  \left(1, \vec{\beta}\right)^\mu.
\ee
We require the conservation equations at order $n^0$. To anticipate the order in the $n$-expansion required for some terms in the stress tensor, we note that, 
\bea
\nabla_A T^{Am} = \left(\partial_\mu + \frac{1}{2} h^{\rho\sigma}h_{\rho\sigma,\mu}\right)T^{\mu m}\\
\nabla_A T^{A\mu} = \bar{\nabla}_\nu T^{\nu\mu}
\eea
where $\bar{\nabla}$'s and $\cdot$'s  are constructed with the $p+1$ dimensional metric, $h_{\mu\nu}$, only.
Since $\Gamma \sim O(n)^{0}$ we require $T^{\mu m},T^{\mu \nu}$ only up to order $n^0$ to get the conservation equations up to $n^0$. We find that all mixed components vanish: $T_{(0)}^{m\mu} = T_{(1)}^{m\mu} = T_{(2)}^{m\mu} = 0$.
For $T_{(0)}$, the only components that appear at order $n^0$ are,
\bea
T_{(0)}^{00} &=& \epsilon,\\
T_{(0)}^{0i} &=& \epsilon\beta^i,\\
T_{(0)}^{ij} &=& \epsilon(\beta^i\beta^j+\gamma^{ij}).\label{eq:T0ij}
\eea
For $T_{(1)}$ we have
\bea
T_{(1)}^{00} &=& O\left(\frac{\eta}{n}\right)\\
T_{(1)}^{0i}& =& O\left(\eta\right)\\
T_{(1)}^{ij} &=& O\left(\eta n\right)
\eea
and so we choose 
\be
\eta = \frac{\bar{\eta}}{n} \label{etascaling}
\ee
and then the only components that appear at order $n^0$ are,
\be
T_{(1)}^{ij} = -\bar{\eta}\sigma^{ij}.
\ee
In order to understand the scaling of $T_{(2)}$, we first need to study how $\mathcal{O}_1$ scales with $n$. Recall that for any finite $d$ the Hawking temperature associated with the event horizon is 
\begin{equation}
\frac{1}{\epsilon ^{1/d} r_0}=\frac{d}{4 \pi T} \quad \Rightarrow \quad  \ln{T}\sim \frac{1}{n}\, \ln \epsilon\,.\label{Temperature}
\end{equation}
Using this we get that 
\begin{align}
\mathcal{O}^{ij}_1&=n^2 \mathcal{R}^{ij}-n^3 \nabla^i \nabla^j \ln{T}+n^3\, \nabla^i \ln{T}\,\nabla^j \ln{T}\nonumber\\
&=n^2 (\mathcal{R}^{ij}- \nabla^i \nabla^j \ln{\epsilon}) +O(n)\nonumber\\
\mathcal{O}^{0j}_1&=O(n)\nonumber\\
\mathcal{O}^{00}_1&=O(n)^{0}\nonumber
\end{align}
Overall we find that
\bea
T^{00}_{(2)} = O\left(c_1,c_2,\frac{c_3}{n},\frac{c_4}{n},\frac{c_4}{n}\right)\\
T^{0i}_{(2)} = O\left(c_1 n,c_2 n, c_3,c_4,c_5\right)\\
T^{ij}_{(2)} = O\left(c_1 n^2, c_2 n^2, c_3\, n, c_4\, n, c_5\, n\right)
\eea
and thus, in order for all the terms above to contribute at leading order, we take the constants $c_i$ to have the following dependence on $n$
\begin{align}
&c_1=\frac{\bar{c}_1}{n^2}\,, c_2=\frac{\bar{c}_2}{n^2}\,\nonumber\\
&c_3=\frac{\bar{c}_3}{n},c_4=\frac{\bar{c}_4}{n},c_5=\frac{\bar{c}_5}{n}\,.\label{cdefs}
\end{align}
For reasons that will be clear in the following section we choose
\begin{align}
&\bar{c}_2=\bar{c}_3=\bar{c}_4=\bar{c}_5=0\, ,\label{cdefs1}
\end{align}
so that only $\mathcal{O}^{ij}_1$ will contribute at leading order
 \bea
T^{ij}_{(2)} =\bar{c}_1 (\mathcal{R}^{ij}- \nabla^i \nabla^j \ln{\epsilon}).
\eea

The conservation equations are then, for the $0$-component,
\bea
\nabla_A T^{A0} = \nabla_A T^{A0}_{(0)} + O(n)^{-1} = \left(\partial_0 + K\right)\epsilon + \nabla_i\left(\epsilon \beta^i\right) + O(n)^{-1} \label{hydro0}
\eea
and for $i$-component, 
\bea
\nabla_A T^{Ai} = \nabla_\mu T^{\mu i}_{(0)} + \nabla_j T^{ji}_{(1)} + \nabla_j T^{ji}_{(2)}  + O(n)^{-1}.\nonumber
\eea
where
\bea
\nabla_\mu T^{\mu i}_{(0)} &=& \gamma^{ij}(\partial_0 + K)(\epsilon \beta_j) + \nabla_j(\epsilon \beta^i\beta^j) + \nabla^i \epsilon\nonumber\\
&&- \epsilon \, \gamma^{ij}( \partial_t \zeta_j +2\beta^l \nabla_{[l}\zeta_{j]}+\frac{1}{2} \nabla_j\gamma_{tt})\,,\label{hydroid0}\\
\nabla_j T^{ji}_{(1)}&=&-\nabla_j\left(\bar{\eta} \nabla^j \beta^i + \bar{\eta}\nabla^i \beta^j + 2\bar{\eta} K^{ij}\right) \,, \label{hydroid1}\\
\nabla_j T^{ji}_{(2)}&=& \nabla_j\left( \bar{c}_1\mathcal{R}^{ij}- \bar{c}_1\nabla^i \nabla^j \ln{\epsilon}\right)\,,\label{hydroid2}
\eea
where the terms proportional to $\zeta$ and $\gamma_{tt}$ come from Christoffel connections.

\subsection{Matching large-d hydrodynamics to $p=2$ large-d gravity\label{sec:matching}}
The aim of this section is to compare the constraint equations \eqref{resultsnozeta0} and \eqref{resultsnozetai} coming from the gravity calculation presented in section \ref{sec:gravity} to the large-D hydrodynamic theory described by \eqref{hydro0}-\eqref{hydroid2}. 

We start by rewriting the gravity equation \eqref{resultsnozeta0} as
\be
(\partial_t +K)a + \nabla_i(\tilde{p}^i-\nabla^i a) = 0\,.
\ee
It is easy to see that it matches exactly with equation \eqref{hydro0} given the following identifications (up to an overall normalisation)
\bea
a = \epsilon\nonumber\\
\tilde{p}^i - \nabla^i a = \epsilon \beta^i\,.\label{identifications}
\eea
We next consider the spatial component of the constraint equations. Using the definitions \eqref{identifications} and taking advantage of identities that hold specifically for  $p=2$, equation \eqref{resultsnozetai} can be brought to the following form 
\bea\label{eq:hydroeq}
(\partial_t +K- \nabla_j\nabla^j )\left(\epsilon\beta_i + \nabla_i \epsilon\right)&=& - \lambda \nabla_i \epsilon - \nabla_j\left(\frac{(\epsilon\beta_i + \nabla_i \epsilon) (\epsilon\beta^j + \nabla^j \epsilon)}{\epsilon}\right)\nonumber\\
 & & + \frac{\epsilon\beta_i}{2}{\cal R} -\nabla_j \left(\epsilon {\cal R}^{j}_{~i}\right)- \epsilon \partial_iK + 2\nabla_m(\epsilon K^m_{~~i})\nonumber\\
 & &+ \epsilon \partial_t \zeta_i + 2 \,\epsilon\,\beta^j\nabla_{[j}\zeta_{i]}+\frac{\epsilon}{2}\nabla_i \gamma_{tt}\nonumber \,.
\eea
This expression contains at most three derivatives. For comparison with the hydro results above let us consider terms with a definite number of derivatives:

\noindent \underline{One-derivative terms:}
\bea
(\partial_t +K) (\epsilon\beta_i ) + \nabla_j \left(\epsilon \beta^j \beta_i\right) +\lambda  \nabla_i \epsilon- \epsilon \partial_t \zeta_i -2\,\epsilon\,\beta^j \nabla_{[j}\zeta_{i]}-\frac{\epsilon}{2}\nabla_i \gamma_{tt}
\eea
which matches \eqref{hydroid0} for the asymptotically $AdS$ case, $\lambda=1$. 

\noindent  \underline{Two-derivative terms: }
\bea
 -\nabla_j\left(\epsilon\nabla_i \beta^j + \epsilon\nabla^j \beta_i+2\epsilon K^j_{~~i}\right)
\eea
which matches \eqref{hydroid1} provided $\bar{\eta} = \epsilon$. From the definition of $\bar{\eta}$ \eqref{etascaling}, using \eqref{Temperature} and the equilibrium relation $\epsilon +p = Ts$, we obtain the dimensionless quantity 
\begin{equation}
\frac{\eta}{s} = \frac{1}{4\pi} + O(n)^{-1}.
\end{equation}

\noindent 
\underline{Three-derivative terms:}
\bea
- \nabla_j\nabla^j \nabla_i \epsilon + \nabla_j \left(\epsilon {\cal R}^{j}_{~i}\right) + \nabla_j\left(\frac{(\nabla_i \epsilon) ( \nabla^j \epsilon)}{\epsilon}\right)
\eea
which matches \eqref{hydroid2} provided $\bar{c}_1= \epsilon$. From this and \eqref{cdefs} we identify the dimensionless combination $\frac{c_1}{s}T = \frac{1}{16\pi^2}$.

Let us now interpret our results in terms of the usual transport coefficients given in equation \eqref{eq:traspcoeffs}. To achieve this we start by making two observations: the first one is that, given \eqref{cdefs} and \eqref{cdefs1}, all transport coefficients scale like $\sim n^{-2}$ and the second one is that they all have the same units as $c_1$. These two points lead us to conclude that 

\begin{equation}
\frac{\lambda_i T}{s}, \frac{\kappa T}{s},\frac{( \eta \tau_\pi) T}{s}= O(n)^0\,. 
\end{equation}
From the matching with hydrodynamics at leading order in $n$, we find that
\begin{align}
&\frac{( \eta \tau_\pi) T}{s}=\frac{1}{16 \pi^2}+ O(n)^{-1}\,, \label{eq:transp_result1}\\
&\frac{\kappa T}{s}=\frac{1}{16 \pi^2}+ O(n)^{-1}\,.\label{eq:transp_result2}
\end{align}
Our result for $\kappa$ matches exactly the result of \cite{Bhattacharyya:2008jc} for $\mathcal{N}=4$ in $D=5$ bulk dimensions, while the result for $\tau_\pi$ matches the universal piece of the formula given in \cite{Haack:2008cp} for $d=3,4,5,6$. To determine $\frac{\lambda_i T}{s}$ one has to go to next to leading order in the matching to hydrodynamics.

It is interesting to point out that if we change the sign of the second term in the definition of $T^{AB}_{(0)}$ \eqref{eq:T0_def}, namely the pressure term, and repeat the calculation of the previous subsection, we find that the only change in the hydrodynamic equations \eqref{hydroid0} is $ \nabla_i \epsilon\to - \nabla_i \epsilon $. This is sufficient for the matching with the large-n gravity calculation in the asymptotically flat case, $\lambda=-1$. Having negative pressure leads to imaginary speed of sound, which matches nicely the fact that the asymptotically flat case suffers from Gregory-Laflamme instabilities, as we will see more explicitly in the next section.

\subsection{Identifying the energy current}
Given the identifications \eqref{identifications}, we can rewrite equation \eqref{resultsnozeta0} as a covariant conservation equation,
\bea
\nabla_A {\cal J}^A &=&  O(n)^{-1},\\
{\cal J}^A &=& \epsilon\, U^A\,, \qquad U^A = \frac{1}{\sqrt{1-\frac{1}{n}\gamma_{tt}-\frac{1}{n}\gamma_{ij}\beta^i\beta^j+ \frac{2}{n}\zeta_i \beta^i}}  \left(1, \vec{\beta}, \vec{0}\right)^A.
\eea
In other words, with a timelike, unit-normed vector $U^A$ we have an associated conserved current ${\cal J}^A$ -- the energy current.
At leading order in $n$, in $a,\tilde{p}_i$ variables,
\bea
{\cal J}^0 &=& a\\
\label{Ji}
{\cal J}^i &=&  \gamma^{ij}\left(\tilde{p}_j- \partial_j a \right)\\
{\cal J}^m &=&  0.
\eea
Note that ${\cal J}_i = O(n)^{-1}$.

\section{Lattices, quasinormal modes and transport\label{sec:transport}}

In this section we consider solutions to the large-$d$ effective equations \eqref{resultsnozeta0} and \eqref{resultsnozetai} corresponding to a monochromatic lattice in two directions with amplitude $A_0$, sourced by the boundary metric deformations,
\be
\gamma_{tt} = \zeta_{i} = 0, \qquad \gamma_{ij} = \left(1 + A_0 \cos{k_L x} \sin{k_L y}\right)\delta_{ij}.
\ee
In addition, we wish to study the transport properties of the system in the presence of this metric deformation.  
In the neutral case, the operators of interest are the energy and heat currents. 
We can extract the corresponding Green's functions by considering infinitesimal sources
$\gamma_{tt}$ and $\zeta_i$ in \eqref{resultsnozetai} and calculating the induced 
energy and heat currents using \eqref{Ji}, see e.g. \cite{Hartnoll:2009sz}. More concretely,  
we will introduce perturbations of definite frequency $\omega$ and momentum $q$, 
\begin{equation}
\label{zeta source}
  	\zeta_i(t, \vec x) = e^{- i \omega t + i\vec q \cdot \vec x} \zeta_i^{(0)}, 
  	\qquad \gamma_{tt}(t, \vec x) = e^{- i \omega t + i\vec q \cdot \vec x} \gamma_{tt}^{(0)},
\end{equation} 
\noindent where $\zeta_i^{(0)}$ and $\gamma_{tt}^{(0)}$ are constants. 
Introducing the linearised sources of the form \eqref{zeta source}, turns on coupled 
perturbations of all the fields as
\begin{equation}
\label{lin pert}
	\delta a = e^{- i \omega t + i\vec q \cdot \vec x} \delta a_{\vec k_L} (x), \qquad 
	\delta \tilde p_i = e^{- i \omega t + i\vec q \cdot \vec x} \delta \tilde p_{i, \vec k_L} (x)
\end{equation}
\noindent where $\delta a_{\vec k_L} (x)$, $\delta \tilde p_{i, \vec k_L} (x)$ are lattice periodic functions. 
Note that due to the inhomogeneity of the background, even spatially homogeneous sources induce 
an inhomogeneous response. Moreover, even though the background is inhomogeneous, linearity of the equations 
allows us to consider fluctuations of the form \eqref{lin pert}. In this sense, $\vec q$ is the 
Bloch pseudo-momentum of the perturbation. This means in particular that in the linearised equations
all functions have the periodicity of the lattice and $q$ appears as a free parameter. 

We will focus on the zero modes of the Green's functions, see e.g. \cite{Donos:2017gej}, by computing the response current
\eqref{Ji} and taking its lattice average. More concretely, after solving the equations of motion for $\delta a_{\vec k_L} (x)$, 
$\delta \tilde p_{i, \vec k_L} (x)$, we compute the perturbed current as
\begin{equation}
	\delta {\cal J}^i  (x) = \gamma^{ij} [ \delta \tilde p_{i, \vec k_L} (x) - (\partial_i + i q_i)  \delta a_{\vec k_L} (x)   ]
\end{equation}
\noindent and take the lattice averages,
\begin{align}
	\delta  J^i &:= \frac{k_L^2}{(2 \pi)^2} \int d^2 x \sqrt{\gamma} \delta  {\cal J}^i  (x), \\
	\delta E &:=   \frac{k_L^2}{(2 \pi)^2} \int d^2 x \sqrt{\gamma} \delta a_{\vec k_L} (x) .
\end{align}
In Appendix \ref{app:zero modes}, we show that this procedure coincides with the definition of the Green's functions zero modes given 
in \cite{Donos:2017gej}.
Having obtained the correlators, the Green's functions then read
\begin{align}
\label{GG}
	\delta J^i (\omega, q)  &=  G^{ij} (\omega, q) \delta \zeta_j^{(0)}, \\
\label{GG2}
	\delta E  &= 2 G(\omega, q) \delta \gamma_{tt}^{(0)}.
\end{align}
The normalisation in \eqref{GG}, \eqref{GG2} is chosen such that the Green's functions satisfy 
\begin{equation}
\label{ward}
	q_i q_j G^{ij}(\omega, q) = \omega^2 G(\omega, q).
\end{equation}
This relation must hold on general grounds \cite{Donos:2017gej}, and we have verified it 
in our numerical calculations. We can understand it directly from the equations of motion 
\eqref{resultsnozetai},  by noting that a finite frequency and momentum 
perturbation with $\delta \gamma_{tt}^{(0)}$ is equivalent to one with 
$\delta \zeta^{(0)}_i = 2 \omega^{-1} q_i$. This shows that all the information 
about transport at finite frequency and momentum is encoded in $G^{ij}(\omega, q)$, so we focus on this correlator below.
The conductivity $\kappa_{ij}(\omega, q)$ can be written as
\begin{equation}
	\kappa^{ij}(\omega, q) = \frac{i}{\omega}  G^{ij} (\omega, q). 
\end{equation}

The poles of the two-point functions are given by the set of QNMs, i.e. non-trivial solutions to the perturbation equations for which the 
sources are zero. The frequencies of these excitations are generically complex, due to the presence of dissipation. 
We will obtain these below, and find agreement between them and the poles of the conductivity. 

For orientation it is useful to recall the results in the absence of a lattice. 
As shown in \cite{Emparan:2016sjk}, the QNM spectrum of the translational invariant theory of pure gravity in the large $d$ limit 
is given entirely by the hydrodynamic modes.  In fact, choosing a static background, $a = 1$, $\tilde p_i = 0$, 
the modes can be written as follows, 
\begin{equation}
	\delta a = e^{- i \omega t + i\vec q \cdot \vec x} \delta a_0, \qquad 
	\delta \tilde p_i = e^{- i \omega t + i\vec q \cdot \vec x} \delta \tilde p_{i0}
\end{equation}
\noindent where $\delta a_0$, $\tilde p_{i0}$ are constants which satisfy relations given by the equations of motion.
The solutions can be classified as
\begin{align}\label{dis p shear}
&{\rm shear}:    &\delta a_0 &= 0,  &\omega &= - i q^2 \\
\label{dis p sound}
&{\rm sound}:  &\delta a_0 &\neq 0,  & \omega &= \pm \sqrt{\lambda}q  - i q^2.
\end{align}
By placing the system in a box, still at non-zero lattice amplitude, we obtain an infinite tower of modes, 
some of which remain gapless and closest to the real axis, i.e. in the Brillouin zone, 
$0 < \vec q < \vec k_L$. The separation between the modes in the Brillouin zone and the higher 
modes is given by integer multiples of $k_L = 2 \pi/L$, so that they all coincide in the limit $L \to \infty$. 
In the presence of a non-trivial background, the order of the equations of motion remains the same, 
so we expect the number of modes to be kept constant when turning on $A_0$. The effect of the lattice 
is then to distort the dispersion relations \eqref{dis p shear}, \eqref{dis p sound}, with the corresponding change in the eigenfunctions
$a_{\vec k_L} (x)$, $\delta \tilde p_{i, \vec k_L} (x)$. 
For moderate $A_0$, the modes in the Brillouin zone 
will remain the closest modes to the real axis and hence will control transport, so we will restrict our attention to these excitations.  

We can also obtain an analytic result for the thermal conductivity in the translationally invariant case. 
Here we restrict ourselves to the AdS case ($\lambda =1$) in which the holographic interpretation 
of our results is clear. 
Considering perturbations with momentum along the $x$ direction, $\vec q = q \hat e_x$, we find 
\begin{align}
	\kappa^{xx} &= \frac{2 q^2 - i \omega}{(- i q + q^2 - i \omega)(i q + q^2 - i \omega)},\\
	\kappa^{yy} &= \frac{1}{q^2 - i \omega},\\ 
	\kappa^{xy} &= 0,\\
	\kappa^{yx} &= 0.
\end{align}
As expected, the poles of the conductivity matrix are located at the QNM frequencies \eqref{dis p shear}, \eqref{dis p sound}. 
More precisely, the longitudinal conductivity $\kappa^{xx}$ is governed by the sound modes, while the transverse conductivity
$\kappa^{yy}$ is controlled by the shear mode. 
The Kramers-Kronig relation dictates that there should be delta functions in the real part of the 
conductivities at the locations where the imaginary parts have poles. We omit this to ease notation. 
We observe that at $q= 0$, both $\kappa^{xx}$ and $\kappa^{yy}$
have a single pole in the imaginary part at $\omega=0$, so that the conductivities have a delta function $\delta (\omega)$ 
in the real part. This delta will be removed by the introduction of the lattice.

\subsection{Perturbation theory in $A_0$}

We first proceed order-by-order in a small $A_0$ expansion at fixed $k_L$. We begin with the lattice solutions themselves which we denote with subscript `$L$',
\be
a(t,x,y) = a_L(x,y), \qquad p_x(t,x,y) = p_L(x,y), \qquad p_y(t,x,y) = p_L(y,x). \label{aplat}
\ee
Note that in \eqref{aplat} we have already imposed a discrete lattice symmetry exchanging $x$ and $y$.
To solve we construct each quantity at order $(A_0)^N$ using an ansatz consisting of products of sines and cosines,
\be
\phi^{(N)}(t,x,y) = \sum_{s=0}^{N}\sum_{m,n = 0}^s c_{mns} s_x^{m} c_x^{m-s} s_y^{n} c_y^{n-s} \label{trigansatz}
\ee
where for compactness we have introduced $s_x \equiv \sin{k_Lx}, c_x \equiv \cos{k_Lx}$, while $c_{mns}$ are constants. 
After solving the equations at each order there are a set of unfixed integration constants. We fix these by demanding that $a_{L}(0,0) = 1$ and $p_{L}(0,0) = 0$; this uniquely specifies the perturbative solution. We have performed this calculation up to and including $N=4$; this will be the required order to extract some of the most relevant physical features of QNMs, as we see shortly in section \ref{pertqnm}. We first provide the explicit expressions for the background up to  $N=1$,
\bea
a_L^{(0)} &=& 1,\\
p_L^{(0)} &=& 0,\\
a_L^{(1)} &=& k_L^2\frac{1-c_xc_y}{2k_L^2+\lambda},\\
p_L^{(1)} &=& k_L^3\frac{s_xc_y}{2k_L^2+\lambda}.
\eea
Note that the linear response piece of this calculation, i.e. the functions at order $(A_0)^1$, diverge in the asymptotically flat case ($\lambda = -1$) at the point $k_L = \pm \frac{1}{\sqrt{2}}$. This corresponds to the Gregory-Laflamme zero mode at unit-wavenumber\footnote{Note our perturbation has a total momentum $\sqrt{k_x^2+k_y^2} = \sqrt{2}k_L$.} \cite{Gregory:1993vy}, 
and thus the onset of a branch of stationary non-uniform black branes.

In the asymptotically AdS case ($\lambda = 1$), we may compute the hydrodynamic variables using \eqref{identifications}, finding zero fluid velocity,
\bea
\epsilon &=& a_L(x,y) + O(A_0)^5,\\
\beta^i &=& O(A_0)^5.
\eea

\subsubsection{Quasinormal modes \label{pertqnm}}

Next we compute QNMs for the perturbative lattice backgrounds. For this we require linear deviations controlled by $\varepsilon$ of the following form,
\bea
a(t,x,y) &=& a_L(x,y) + \varepsilon e^{-i \omega t + i q x} \sum_{i=0}\delta a^{(i)}(x,y) (A_0)^i\\
p_x(t,x,y) &=& p_L(x,y) +\varepsilon e^{-i \omega t + i q x} \sum_{i=0} \delta p_x^{(i)}(x,y) (A_0)^i\\
p_y(t,x,y) &=& p_L(y,x) +\varepsilon e^{-i \omega t + i q x} \sum_{i=0}\delta p_y^{(i)}(x,y) (A_0)^i \label{qnmansatz}
\eea
where $a_{L}(x,y)$ and $p_{L}(x,y)$ are obtained earlier, $\delta a^{(0)}(x,y) , \delta p_x^{(0)}(x,y) , \delta p_y^{(0)}(x,y) $ are taken to be constant, and
\be
\omega = \sum_{i=0}\omega^{(i)}(A_0)^i,\qquad q = \sum_{i=0}q^{(i)}(A_0)^i.\label{omegaqpertexpansion}
\ee
Thus we are considering a double expansion in small $A_0$ and $\varepsilon$, but keeping up to linear order in $\varepsilon$ only. Note that we have specialised to perturbations with momentum in the $x$ direction without loss of generality. There are two mode classes, here characterised by whether $\delta p_y^{(0)}= 0$ or $\delta a^{(0)} = 0$, which we label as \emph{sound} and \emph{shear} modes respectively.
We can focus on different regimes of $\omega, q$ by choosing the leading behaviour of \eqref{omegaqpertexpansion}. For $\omega \sim q \sim (A_0)^0$ both sound and shear modes obey standard hydrodynamic dispersion relations with quadratic corrections due to the lattice, i.e.
\bea
\omega =& -i q^2 + O(A_0)^2\qquad &\text{(shear)}\\
\omega =&\pm \sqrt{\lambda} q -i q^2 + O(A_0)^2\qquad &\text{(sound)}.
\eea
Note in the asymptotically flat case ($\lambda = -1$) this includes the Gregory-Laflamme zero mode at $q = \pm 1$.
We now turn to the regime $\omega \sim q \sim (A_0)^2$ where there is a collision which marks the boundary of purely diffusive mode behaviour. Hence we choose $\omega^{(0)} = \omega^{(1)} = q^{(0)} = q^{(1)} = 0$ from the outset. We continue to solve the equations order-by-order in $A_0$, adopting the same trigonometric ansatz for the fluctuations as we did for the lattice functions, \eqref{trigansatz}. 

For the shear mode, $\delta a^{(0)} = 0$, we here present the leading order perturbation functions,
\bea
\delta a^{(1)}(x,y)  &=& \delta p_y^{(0)} \frac{\lambda k_L}{(2 k_L^2 + \lambda)^2}c_x s_y\\
\delta p_x^{(1)}(x,y)  &=& \delta p_y^{(0)} \frac{1 + \lambda k_L^2+ 2k_L^4 }{2(2 k_L^2 + \lambda)^2} s_x s_y \\
\delta p_y^{(1)}(x,y)  &=&\delta p_y^{(0)} \frac{1 + 5 \lambda k_L^2 + 2 k_L^4}{2(2 k_L^2 + \lambda)^2} (c_x c_y - 1)
\eea
and we have calculated up to $N=4$, finding the dispersion relation
\be
\omega = -i \Gamma - i q^2 + O(A_0)^5\qquad \text{(shear)}\label{pertshear}
\ee
where the gap is set by
\bea
\Gamma &\equiv& \frac{k_L^2(1+2\lambda k_L^2 + 2 k_L^4)}{4(\lambda + 2 k_L^2)^2}(A_0)^2 + \label{Gammadef}\\
&&\frac{k_L^2(c_0 \lambda + c_1 k_L^2 + c_2\lambda k_L^4 + c_3 k_L^6 + c_4 \lambda k_L^8 + c_5 k_L^{10} + c_6 \lambda k_L^{12} + c_7 k_L^{14}+c_8\lambda k_L^{16}  + c_9 k_L^{18}) }{128(\lambda + 2k_L^2)^5(\lambda + 4k_L^2)^2(\lambda + 8k_L^2)^2}(A_0)^4 \nonumber
\eea
where the coefficients are $c_0 \equiv 15, c_1 \equiv 386, c_2 \equiv 4323, c_3 \equiv 28622, c_4 \equiv 126786, c_5 \equiv 383424, c_6 \equiv 772616, c_7 \equiv 995616, c_8 \equiv 744832, c_9 \equiv 249344$.

For the sound modes, $\delta p_y^{(0)} = 0$, we here present the leading order perturbation functions,
\bea
\delta p_x^{(0)} &=& \frac{\omega^{(2)}}{q^{(2)}}\delta a^{(0)}\\
\delta a^{(1)} &=& \delta a^{(0)}\frac{k_L}{\lambda + 2 k_L^2} \left(k_L(1-c_xc_y) + \frac{s_x c_y \lambda}{\lambda + 2 k_L^2}\frac{ \omega^{(2)}}{q^{(2)}}\right) \\\
\delta p_x^{(1)} &=&\delta a^{(0)}\frac{1}{\lambda + 2 k_L^2}\left(s_x c_y k_L^3 + \frac{(c_xc_y-1)(1 + 5 \lambda k_L^2 + 2 k_L^4)}{2(\lambda + 2 k_L^2)}\frac{ \omega^{(2)}}{q^{(2)}}\right)\\\
\delta p_y^{(1)} &=&\delta a^{(0)} \frac{s_y}{\lambda + 2 k_L^2}\left(c_x k_L^3 + s_x\frac{1 + \lambda  k_L^2 + 2  k_L^4}{2(\lambda + 2 k_L^2)}\frac{ \omega^{(2)}}{q^{(2)}}\right)
\eea
where $\omega^{(2)}$ and $q^{(2)}$ are related by the dispersion relation which we have calculated up to order $A_0^6$,
\be
i\omega(-i\omega + \Gamma +\alpha_2 q^2) - \alpha_3 q^2 = O(A_0)^7 \qquad \text{(sound)}\label{pertsound}
\ee
where $\Gamma$ is given in \eqref{Gammadef} and
\bea
\alpha_2 &=&2, \label{alpha2def}\\
\alpha_3 &=&  \lambda - \frac{1 + 2 \lambda k_L^2 + 5  k_L^4 + 2 \lambda k_L^6}{8(\lambda + 2 k_L^2)^3} A_0^2.\label{alpha3def}
\eea
This dispersion relation describes two modes, and a collision between them which marks a transition from purely diffusive to standard hydrodynamic behaviour at $\omega \sim q \sim A_0^2$. More concretely, at $q=0$ from \eqref{pertsound} we have a gapless mode, $\omega(0) = 0$, and a gapped mode, $\omega(0) = -i \Gamma$, i.e the same gap as the shear mode (indeed at $q=0$ these modes are degenerate). Increasing $q$ we find a collision at
\bea
\omega &=& -i\frac{\Gamma}{2}\left(1+\frac{\alpha_2}{4\alpha_3}\Gamma\right)+ O(A_0)^5,\\
q &=& \pm\frac{\Gamma}{2\sqrt{\alpha_3}}\left(1+\frac{\alpha_2}{4\alpha_3}\Gamma \right) + O(A_0)^5.
\eea
These results are captured by our numerical analysis in section \ref{sec:numerics}.

\subsubsection{Conductivity}
For the conductivity we once again consider the interesting regime $\omega \sim q \sim (A_0)^2$. Working to order $(A_0)^4$ in the lattice background and the linear response calculation completely determines the conductivity to order $(A_0)^0$, (here we work exclusively in the AdS case, $\lambda = 1$)
\bea
\kappa^{xx} &=& \frac{i \omega \alpha_1}{i\omega(-i\omega + \Gamma + q^2\alpha_2) - q^2 \alpha_3} + O(A_0)^1\label{pertkappaxx}\\
\kappa^{xy} &=& O(A_0)^1\\
\kappa^{yx} &=& O(A_0)^1\\
\kappa^{yy} &=& \frac{i \alpha_1}{\omega + i \Gamma + i q^2}+ O(A_0)^1\label{pertkappayy}
\eea
where $\alpha_2,\alpha_3,\Gamma$ are given in \eqref{alpha2def}, \eqref{alpha3def} and \eqref{Gammadef} respectively, while
\be
\alpha_1 = 1+\frac{k_L^2}{1+2k_L^2}A_0 - \frac{1+30 k_L^2 + 248 k_L^4 + 790 k_L^6 + 1044 k_L^8 + 368 k_L^{10}}{8(1+2k_L^2)^3(1+ 12k_L^2 + 32 k_L^4)} A_0^2.
\ee
We note that the derived expression \eqref{pertkappaxx} exhibits the full sound-channel pole structure found earlier, \eqref{pertsound}, and matches the form of the phenomenological model of momentum relaxation in \cite{Davison:2014lua}. Similarly, \eqref{pertkappayy} exhibits the shear-channel pole structure found earlier, \eqref{pertshear}.

\subsection{Numerical results\label{sec:numerics}}

\begin{figure}[th]
\centering
\includegraphics[width=0.75 \linewidth]{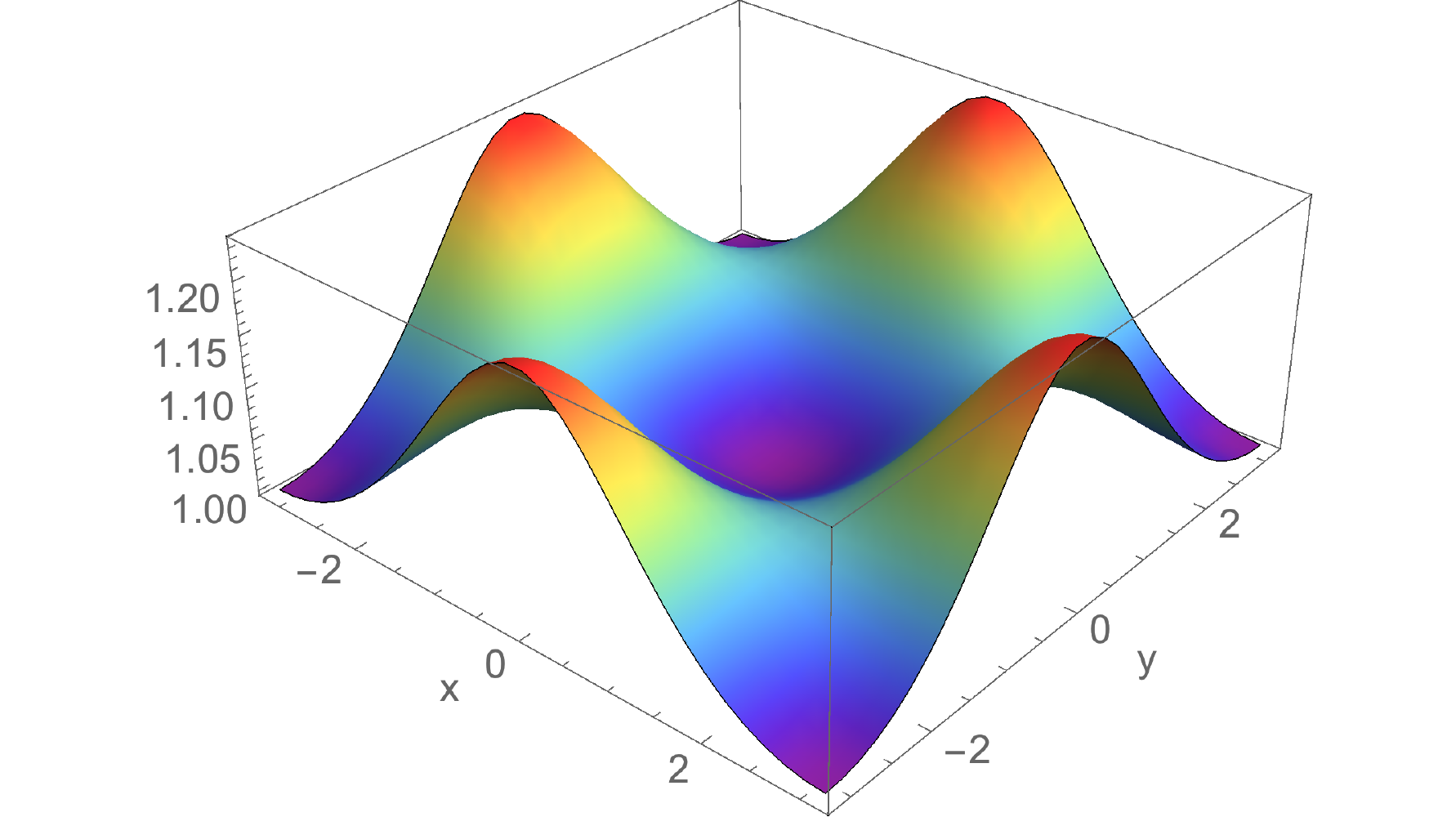}
\caption{Profile of the lattice energy density $a_L(x,y)$ for a lattice \eqref{lattice def} with $A_0 = 0.3$ in the interval $[-\pi, \pi]^2$.}%
\label{fig:aBckg}
\end{figure}

Let us now describe our numerical results.
We first construct the background solutions, taking $\gamma_{tt} = 0$, $\zeta = 0$ and varying the 
amplitude $A_0$ of our monochromatic lattice. In order to obtain a unique solution, we must fix the 
value of $a_L$ in one corner of the lattice. However, this value can be 
rescaled away due to the symmetry of the equations of motion at $\zeta = 0$.
We then fix $a_{corner} = 1$ throughout without loss of generality. 
We solve the numerical problem by discretising the set of PDEs on a homogeneous
grid with periodic boundary conditions, and find the non-linear solutions by relaxation.
We plot the so obtained background profiles in Fig \ref{fig:aBckg}. In all our background 
solutions we have $\tilde p_i = \partial_i a$, so there is no background current. In all 
our numerical calculations we have set $k_L = 1$. 

\subsubsection{Quasinormal modes}

We solve for the QNM numerically by discretising the system of linear PDEs and solving the resulting matrix 
eigenvalue problem. 
It should be noted that we do not require the perturbed energy density to be fixed, but only 
demand periodicity of all the fluctuations. This reproduces correctly the spectrum of the 
$A_0 = 0$ case. 

\begin{figure}[th]
\centering
\includegraphics[width=0.4 \linewidth]{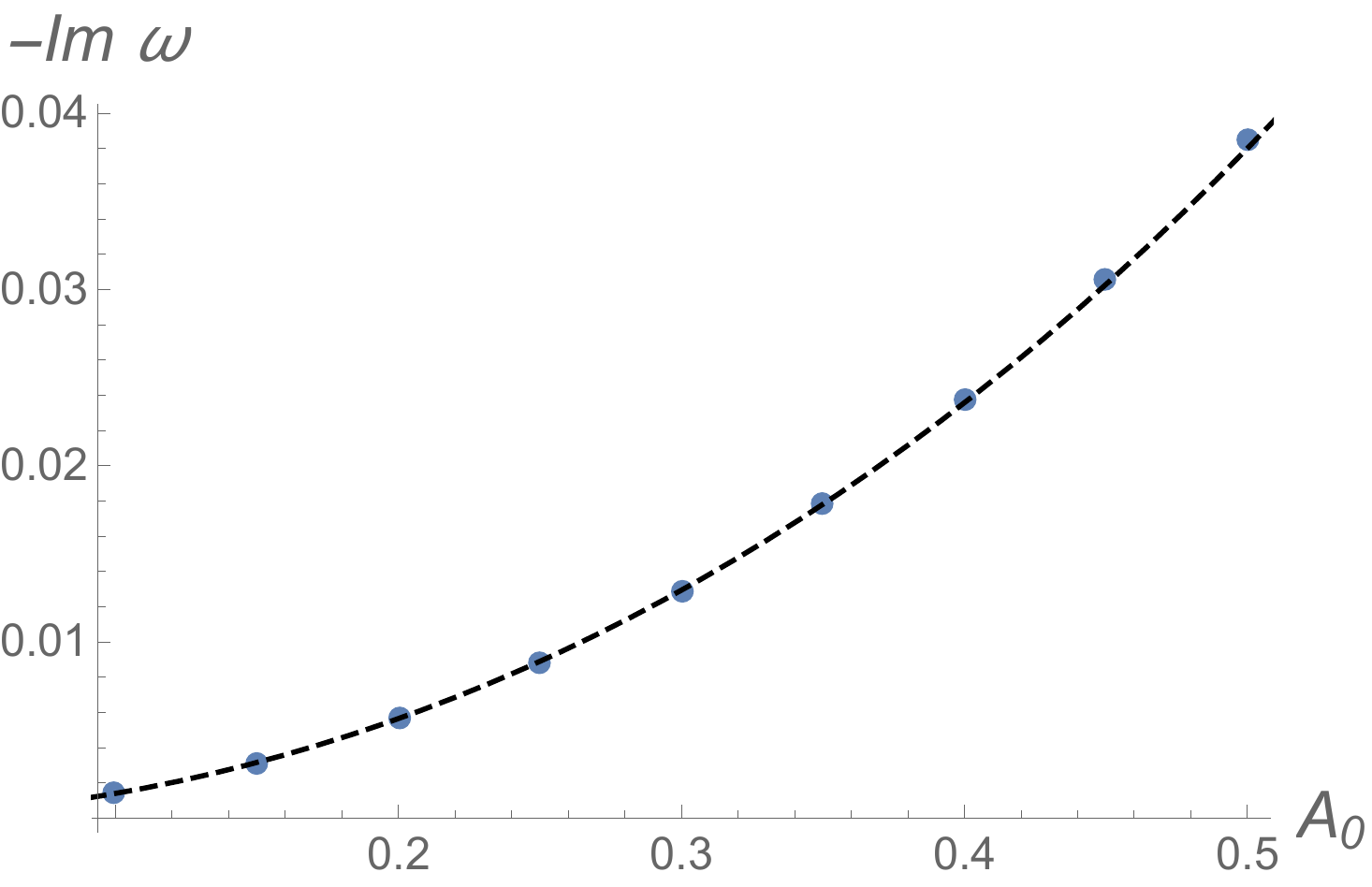}
\caption{Imaginary part of the smallest QNM for varying amplitude at $q=0$. The dashed line shows our perturbative result. }
\label{fig:QNM_q=0}
\end{figure}

We discuss first the $q = 0$ modes. At $A_0=0$, we have 3 zero modes sitting at the origin of the complex $\omega$ plane. 
Increasing $A_0$, we find that one of these remains at $\omega = 0$, while the other two descend together down the 
imaginary axis in excellent agreement with our perturbative calculation \footnote{This degeneracy can be uplifted by considering non-square lattices, e.g. 
$\gamma_{ij} = \delta_{ij}(1 + \cos (n_x x) \cos (n_y y))$ with $n_x \neq n_y$.}. 
We plot our numerical results for the dependence of the imaginary part of the frequency of this double pole in Fig \ref{fig:QNM_q=0}. 
We know from our perturbative calculation that in this regime the conductivity is given by a Drude peak 
with a width determined by the inverse of the imaginary frequency of the double pole. 
We will confirm this numerically in Sec. \ref{sec:kappa}.

Let us now discuss our results at finite $q$. For concreteness, we choose $\vec q$ to be aligned in the $x$ direction. 
Once again, our results coincide with the perturbative calculation \eqref{kappaintro}. In particular, we observe that 
increasing $q$ from $q= 0$ with $A_0$ fixed, the zero mode and one of the double poles move 
towards each other along the imaginary axis, until they collide and move off to the complex plane at $q = q_c$. Since at large 
$q$ we can ignore the effect of the lattice, we identify these modes with the sound modes of the translationally invariant theory. 
On the other hand, the third mode is gapped and remains on the imaginary axis as $q$ increases. Because of this property, 
we identify it with the shear mode. We have checked that for $q_c \ll  q < 1$, 
the dispersion relations coincide with the translational invariant ones, up to small gaps.

For values of $q < q_c$, i.e. before the collision, we find by fitting our numerical data that the lowest QNMs have dispersion relations 
\begin{equation}
\label{disp rel num 1}
	\omega_- = - i D_- q^2, \qquad \omega_+ = - i ( \tau^{-1} - D_+ q^2 ).
\end{equation}
\noindent Note that the lowest QNM is gapless and has a quadratic dispersion relation, in agreement with the prediction of 
\cite{Donos:2017gej}. Moreover, we shall see in section \ref{sec:kappa} that the DC thermal conductivity satisfies precisely 
the Einstein relation derived in that reference.
The dispersion relation for the shear mode is also quadratic,
\begin{equation}
\label{disp rel num 2}
	\omega_{ {\rm shear}} = - i ( \tau^{-1} + D_s q^2 ).
\end{equation}
\noindent Note that $\tau$ is the characteristic time at $A_0$ given by the inverse of the imaginary 
frequency in Fig. \ref{fig:QNM_q=0}. 

\begin{figure}[th]
\includegraphics[width=1.1 \linewidth]{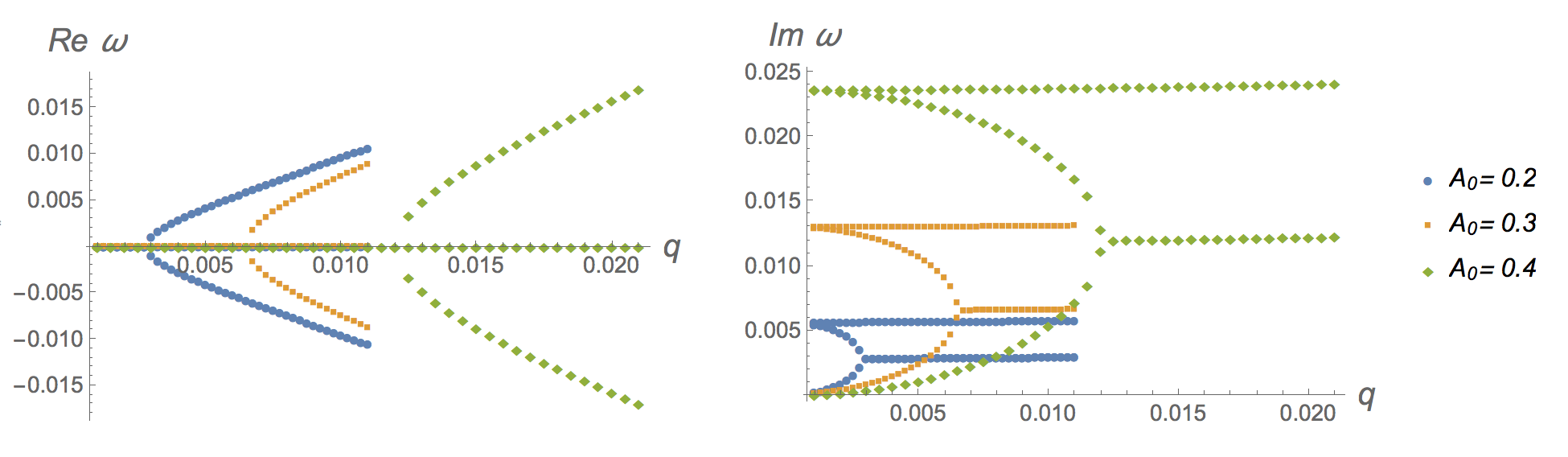}
\caption{Dispersion relation for the lowest modes in the presence of a lattice with $A_0 = 0.2, 0.3, 0.4$. As we increase $q$, 
two modes move towards each other on the imaginary axis, until they collide and acquire a non-vanishing real part. 
The third mode remains on the imaginary axis. After the collision, the curves depicting the dispersion relations 
become essentially featureless, so we do not display the data points for large values of $q$ to ease readability of 
the figures.}%
\label{fig:disp_rel}
\end{figure}

We plot our numerical results for the dispersion relations in the presence of the lattice in Fig. \ref{fig:disp_rel}, 
and provide the values of the fit parameters in Table \ref{table fits}. We find good agreement 
between the perturbative and numerical results. 
We emphasise that the sound modes remain closest to the real axis after and before the collision, 
so they dominate transport. 

\begin{table}
\centering
\begin{tabular}{ |c|c|c|c|c|c| } 
 \hline
                     &  $1/\tau$   &  $D_-$  &  $D_+$ & $q_c$   & $D_s$  \\ 
  \hline
 $A_0 = 0.2$ &  $5.5 \cdot 10^{-3}$   & 182.2  & 182.7 & $2.7 \cdot 10^{-3}$    & 1.0 \\ 
 \hline
  $A_0 = 0.3$ & $1.3 \cdot 10^{-2}$  &  77.3  &  75.5 & $6.5 \cdot 10^{-3}$  & 1.0 \\ 
 \hline
   $A_0 = 0.4$ & $2.3 \cdot 10^{-2}$   &  41.9  & 40.0 & $1.2 \cdot 10^{-2}$  & 1.0  \\ 
 \hline
\end{tabular}
\caption{Parameters of the dispersion relations of the lowest QNMs}
\label{table fits}
\end{table}

The behaviour of the sound modes can be adequately described by the denominator of the 
perturbative expression \eqref{kappaintro}. In particular, this formula accounts for their collision 
and motion off to the complex plane. Neglecting higher order corrections in $A_0$ and $q$,
we can identify the various constants in the dispersion relations \eqref{disp rel num 1}, \eqref{disp rel num 2} as
\begin{equation}
\label{identify param}
	\tau^{-1} = \Gamma, \qquad D_- = \frac{\alpha_3}{\Gamma} , \qquad  D_+ = \frac{\alpha_3}{\Gamma} - \alpha_2.
\end{equation}
The only constant left to fix is $\alpha_1$, which can be obtained from the $\omega = q =0$ of the conductivity.

\subsubsection{Conductivity}
\label{sec:kappa}

We find the conductivity numerically by solving the linear PDEs by discretisation and inversion of the resulting 
matrix problem. As in the computation of the QNMs, we do not fix the energy density of the perturbations, 
but only require all the eigenfunctions to have the appropriate periodicity of the lattice. 

\begin{figure}[th]
\centering
\includegraphics[width=1. \linewidth]{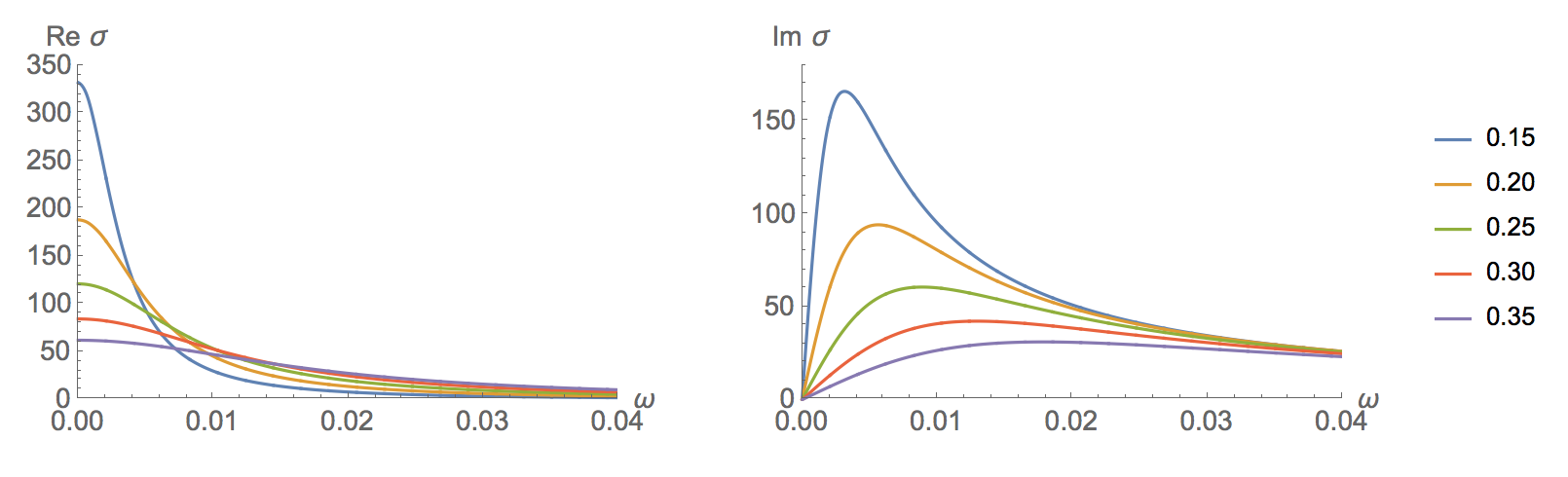}
\caption{Conductivity for $\vec q = 0$. The colour coding corresponds to different values of the lattice amplitude $A_0$.}%
\label{fig:Drude}
\end{figure}

Let us first discuss the results for $\vec q = 0$. For concreteness, we consider the source to have a non-vanishing $x$ component
$\zeta_x^{(0)}$ in \eqref{zeta source}. After averaging, the conductivity on the orthogonal direction vanishes $\kappa^{xy} = 0$, so we will only refer to the 
parallel component $ \kappa := \kappa^{xx}$ in what follows. We obtain the expected Drude peaks, which become less localised as 
we increase the lattice amplitude, see Fig \ref{fig:Drude}. More concretely, the conductivity can be well approximated by the Drude formula,
\begin{equation}
	\kappa(\omega)  = \frac{\kappa_{DC}}{1 - i \omega \tau}.
\end{equation}

\noindent Fitting our numerics to the Drude formula we can extract the parameters $\kappa_{DC}$ and $\tau$. This calculation is in excellent agreement with the one obtained from the QNM in Fig \ref{fig:QNM_q=0} and the perturbative results. 
In particular, the conductivity fit coincides with the value extracted from the QNM within $2 \%$. 
This also allows us to identify the constant $\alpha_1$ in \eqref{kappaintro} 
\begin{equation}
\label{alpha1}
	\alpha_1 = \kappa_{DC}/\tau.
\end{equation}
We show our results for $\alpha_1$ and the remaining fit parameters in Table \ref{table fits alpha}. 

As derived in \cite{Donos:2017gej}, the value of $\kappa_{DC}$ is related to the diffusion 
constant $D_-$ via 
\begin{equation}
\label{eins rel}
	\kappa_{DC} = \chi(q = 0) D_- 
\end{equation}
\noindent where $\chi$ is the charge-charge susceptibility 
\begin{equation}
	\chi(q) = \lim_{\omega \to 0 + i 0} G(\omega, q).
\end{equation}
Note that $\chi(q)$ cannot be obtained directly from the conductivity since the relation 
\eqref{ward} becomes singular at $\omega=0$. Since both $\kappa_{DC} $ and $D_-$ are of the same 
order in perturbation theory, we only need to compute $\chi(0)$ to leading order in order to verify 
\eqref{eins rel}. We find
\begin{equation}
	\chi(0) = 1 + O(A_0^2),
\end{equation}
\noindent which shows that \eqref{eins rel} holds for the perturbative expressions. 
We show our results for $\chi(0)$ in Table \ref{table fits alpha}. We can check that 
\eqref{eins rel} holds up to $3 \%$ in our numerics.

\begin{figure}[th]
\centering
\includegraphics[width=1. \linewidth]{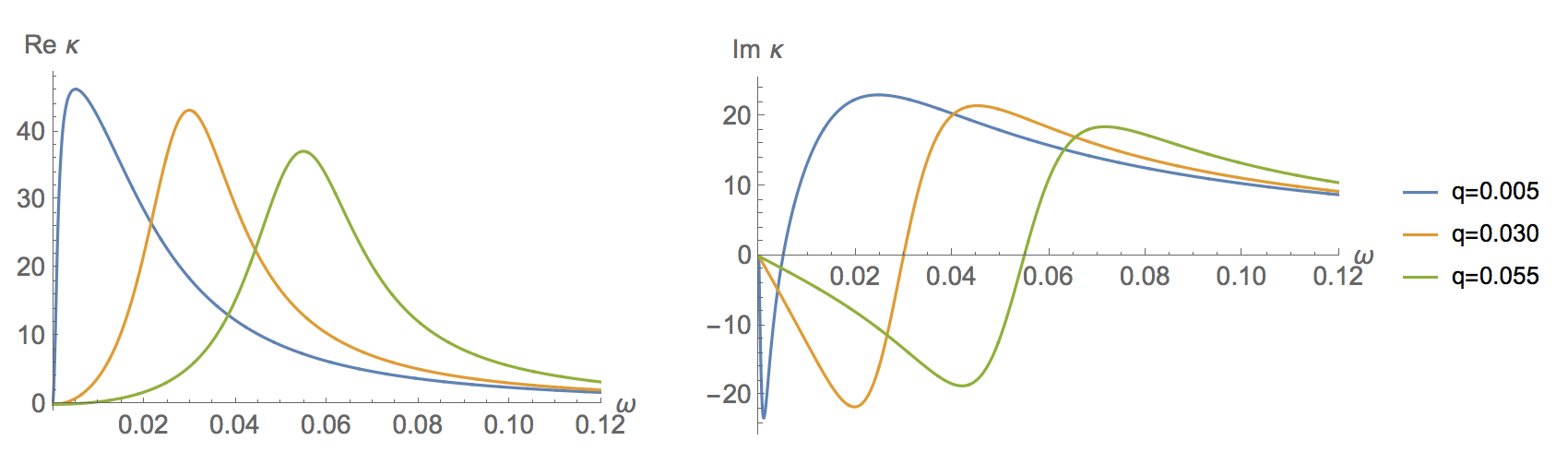}
\caption{Longitudinal conductivity for $A_0 = 0.4$ and varying $q$.}%
\label{fig:resonances}
\end{figure}

We now introduce non-vanishing momentum along the $x$ direction, $\vec q = q \hat e_x$. Considering 
a deformation along $x$ only, $\zeta_y = 0$, it turns out that $\kappa^{xy} =\kappa^{yx}= 0$ as in the $q = 0$ case. 
We plot our results for $\kappa(\omega , q)$ in Fig \ref{fig:resonances}. We distinguish 
two regimes, which are related to the behaviour of the QNMs. At small momentum, $q < q_c$, 
all QNMs lie on the imaginary axis, and ${\rm Re} \kappa$ rises quickly with $\omega$ until it reaches a 
maximum value at $\omega = \sqrt{\alpha_3} q$ and then decays. At large momentum $q > q_c$, 
the sound modes can propagate, so we observe resonances  at $\omega = \sqrt{\alpha_3} q$. 

As stated in the introduction, our results for the conductivity at small $\omega$ and $q$ can 
be described by the perturbative result \eqref{kappaintro}, neglecting higher order corrections in $A_0$. 
To illustrate this, we proceed as follows. 

\begin{figure}[th]
\centering
\includegraphics[width=1. \linewidth]{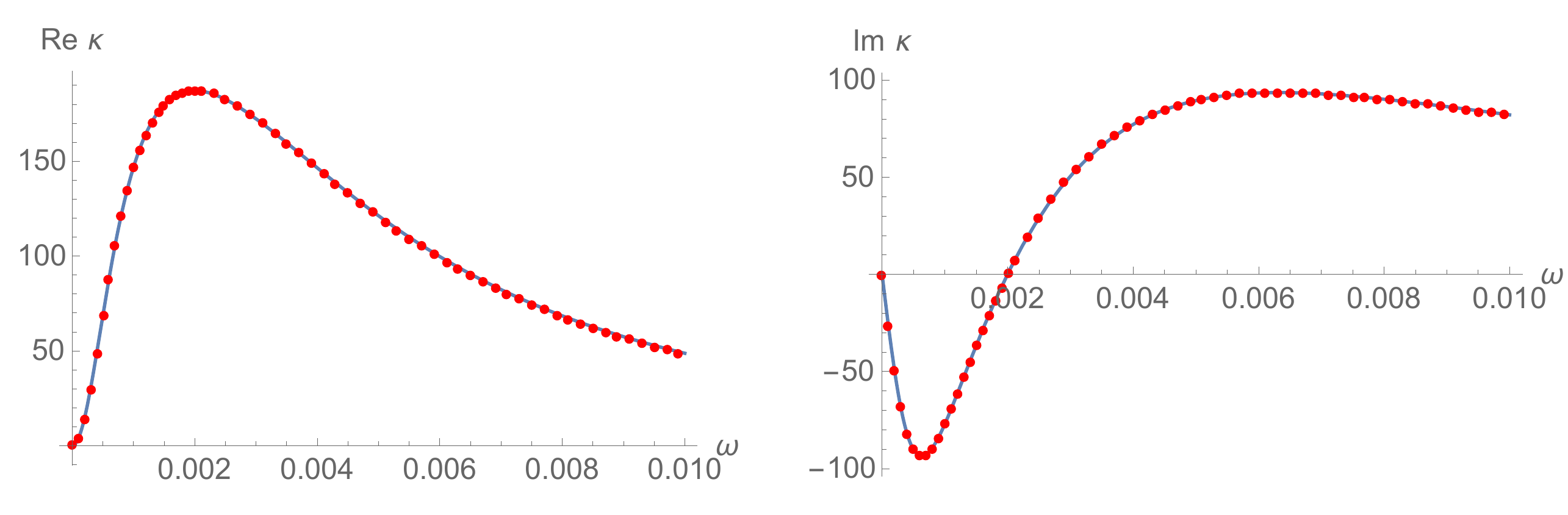}
\caption{Conductivity for $A_0 = 0.2$ and $q=2 \cdot 10^{-3}$. The points correspond to the numerical data, while the 
solid line is obtained from a fit.}%
\label{fig:kappa fit}
\end{figure}

First, we extract the parameters $\Gamma$, $\alpha_i$ from the conductivity data produced by our numerics. 
We obtain $\Gamma$ and $\alpha_1$ from the Drude peaks with $q = 0$, and $\alpha_2$, $\alpha_3$ from the location 
and height of the maximum of $Re \kappa (\omega, q)$ in the $q \neq 0$ data. 
Having done this, we can compare the curves resulting from \eqref{kappaintro} with the so obtained parameters
and the numerical data. We have found excellent agreement in this comparison, see Fig \ref{fig:kappa fit} for an example.
We show our results for the parameter fits in Table \ref{table fits alpha} \\

\begin{table}[h]
\centering
\begin{tabular}{ |c|c|c|c|c|c|c| } 
 \hline
                                      &  $1/\tau$   &  $\alpha_1$ &  $\alpha_2$ & $\alpha_3$  & $\chi(0)$    \\ 
  \hline
 $A_0 = 0.2$ &  $5.69 \cdot 10^{-3}$   & 1.066  & 2.00 & 0.998  & 1.06\\ 
 \hline
  $A_0 = 0.3$ & $1.29 \cdot 10^{-2}$  &  1.08  &  1.96 & 0.995  & 1.09\\ 
 \hline
   $A_0 = 0.4$ & $2.36 \cdot 10^{-2}$   &  1.10  & 1.94 & 0.991  &  1.12\\ 
 \hline
\end{tabular}
\caption{Fit parameters for the conductivity in expression \eqref{kappaintro}
and charge-charge susceptibility.}
\label{table fits alpha}
\end{table}

Comparing these results with the ones obtained from the QNMs (see Table \ref{table fits}) using 
the identifications \eqref{identify param}, we find good agreement. 
In particular, for 
$A_0 = 0.2$ the results differ by $5\% $, while for $A_0 = 0.4$ 
this is reduced to  less than $1\% $. The improvement of this match is due to the fact 
that the shear mode, not accounted for in the formula \eqref{kappaintro},
becomes more separated from the real axis at larger $A_0$. 

\section{Discussion\label{sec:discussion}}
In this paper we generalised the large-$d$ effective near horizon equations to include inhomogeneous and time dependent metric deformations at the boundary of AdS. They are given in \eqref{resultsnozeta0} and \eqref{resultsnozetai}. Our class of metric deformations, i.e. \eqref{bdymet}, do not disturb the general form of the near horizon equations. Instead, the deformations lead to a nontrivial background metric on which the effective theory lives, and adds source terms to the equations. Interpreted as the effective theory corresponding to the light ($\omega \sim O(d)^0$) quasinormal mode sector, our class of metrics correspond to ambient deformations that do not excite any of the heavier degrees of freedom. We have extended these results also to asymptotically flat spacetimes. 

A second key result of this work is the identification of the covariant equations with the conservation of the second order truncated Landau-frame hydrodynamic stress tensor on the nontrivial background. This includes terms at second order which depend on the Ricci curvature of the background metric. Given the agreement between the truncated hydrodynamic theory at large $d$ and the gravitational equations, our result could be interpreted as a matching calculation from the microscopic theory (gravity) to the effective theory (hydrodynamics), resulting in a determination of the transport coefficients (detailed in section \ref{sec:matching}). With this information in hand we could have used a much more restricted ansatz in order to tease out the transport coefficients. Indeed, our ansatz was somewhat restricted because we focused only on non-trivial dynamics in $p=2$ spatial dimensions, however now that we have determined the transport coefficients, the truncated hydrodynamic equations can be used to obtain deformed equations at any $p$. 

An interesting technical detail we encountered was the form of the Landau frame hydrodynamic stress tensor at second order. The only term that contributes at leading order in large $d$ is ${\cal O}_1^{ab}$ (defined in \eqref{Odefs}), with other terms having subdominant scalings. In hydrodynamics, as a derivative expansion, one has the freedom to use lower order equations to reshuffle higher order terms. In effective field theory language there is a redundancy, and in the literature this is traditionally fixed in a way that eliminates ${\cal O}_1^{ab}$ in favour of a different combination of terms. However, we found that large $d$ gravity prefers to keep ${\cal O}_1^{ab}$, in the sense that the equations follow exactly from this truncated stress tensor without invoking any derivative expansion. In other words, had we used the traditional, BRSSS form of the second order stress tensor, the relationship between truncated Landau frame hydrodynamics and the gravitational equations would only be approximate in the derivative expansion.

In light of this detail, it is interesting to ask in a broader context whether there is anything special about the second order-truncated hydrodynamic theory containing only ${\cal O}_1^{ab}$, as selected by gravity. We showed that it is a well behaved theory at large $d$ (in the sense that it reduces to the Einstein equations), but it would be interesting to understand the properties of this theory further at finite $d$. On a related note, it would be interesting to understand if there is an exact relation between truncated hydrodynamics and the gravitational equations at subleading orders in the $1/d$ Taylor expansion.

Finally we can discuss applications of our equations, \eqref{resultsnozeta0} and \eqref{resultsnozetai}. For illustration in this paper we have chosen to focus on constructing 2d monochromatic lattices and their perturbation spectra and linear response properties in section \ref{sec:transport}. In particular, we found that the equations could be solved analytically in small lattice amplitude, $A_0$, both for the backgrounds, the QNM spectrum and for the conductivity. We also gave details on calculating these quantities numerically at finite $A_0$.
There are many open questions to which these equations could be applied. 
For example, it would be interesting to investigate if a coherent/incoherent transition can occur at large D triggered by the collision 
of the modes in the Brillouin zone with modes outside of it. 
In addition, provided the knowledge of the gravitational counterterms, our formalism allows for the computation of the 
Love numbers of AdS black branes, recently performed for $D=4, 5$ in \cite{Emparan:2017qxd}. 
We considered static configurations, but we emphasise that the equations allow for full time dependence, which for example, could be used to drive the system with a time dependent source to investigate statistically steady turbulence. 

\acknowledgments
It is a pleasure to thank Alexander Krikun and Jorge Rocha for interesting discussions, and especially Roberto Emparan for his comments on an earlier version of this manuscript. The work of TA is supported by the ERC Advanced Grant GravBHs-692951. He also acknowledges the partial support of the Newton-Picarte Grant 20140053. CP is supported by the STFC Consolidated Grant ST/P000371/1. BW is supported by the NCCR under grant number 51NF40-141869 `The Mathematics of Physics' (SwissMAP).

\appendix

\section{Zero modes of the Green's functions}
\label{app:zero modes}

In this appendix we show that our computation of the Green's functions zero modes 
corresponds to the definition given in \cite{Donos:2017gej}. Our starting point is the definition 
of a Green's function $G_{AB}$, which gives the linear response of the operator $A$ in the presence of the source $B$. In position 
space we write
\begin{equation}
\label{def G}
	\delta A (t, \vec x) = \int dt dx' G_{AB}(t-t',\vec x, \vec x') \delta B (t',\vec x').
\end{equation}
We solve the equations of motion by introducing the expansions
\begin{equation}
	\delta B (t, \vec x) = e^{- i \omega t + i \vec q \cdot \vec x} \delta B^{(0)}, 
	\qquad \delta A (t, \vec x) = e^{- i \omega t + i \vec q \cdot \vec x} \delta \hat A_{\vec k_L}(\vec x),
\end{equation}
\noindent where $\delta B^{(0)}$ is a constant and $\delta \hat A_{\vec k_L}(\vec x)$ is a lattice-periodic function.
Moreover, we can freely Fourier transform in the time domain due to time-translational invariance, which 
allows us to write
\begin{equation}
	G_{AB}(t-t',\vec x, \vec x') = \int d \omega e^{- i \omega (t-t')} G_{AB}(\omega , \vec x, \vec x').
\end{equation}
Inserting these expansions in the definition \eqref{def G}, and performing the usual manipulations with the time integrals, we arrive at
\begin{equation}
	\delta \hat A_{\vec k_L}(\vec x) = \int dx' e^{i \vec q \cdot (\vec x- \vec x')} G_{AB}(\omega , \vec x, \vec x') \delta B^{(0)}.
\end{equation}
Taking the lattice average we arrive at
\begin{equation}
	\delta \hat A_{\vec k_L}( \vec x) = \delta B^{(0)}  \frac{k_L^2}{(2 \pi)^2} 
	\int d x   \int dx' e^{i \vec q \cdot (\vec x- \vec x')}  G_{AB}(\omega , \vec x, \vec x').
\end{equation}
Up to the constant factor $\delta B^{(0)}$ which we are normalising to one, this coincides with the 
definition of the zero mode of the two point function, equation (2.11) in \cite{Donos:2017gej}.

\bibliographystyle{JHEP-2}
\bibliography{larged}{}

\end{document}